\documentclass[12pt]{article}
\usepackage{amssymb,latexsym}
\usepackage[dvips]{graphicx}
\def\labelmark{}
\def\void{}

{\ifx\void\labelname\def\junk{\end{displaymath}}
\else\def\junk{\end{eqnarray}}\fi\junk\labelmark\def\labelname{}}

\newcommand{\bra}{\begin{array}}
\newcommand{\era}{\end{array}}
\newcommand{\beq}{\begin{equation}}
\newcommand{\eeq}{\end{equation}}
\newcommand{\bqn}{\begin{eqnarray}}
\newcommand{\eqn}{\end{eqnarray}}

\font\mybb=msbm10  at 12pt
\def\bb#1{\hbox{\mybb#1}}

\font\mybbi=msbm10  at 9pt
\def\bbi#1{\hbox{\mybbi#1}}

\def\BC{\bb C}
\def\_\BC{\bbi C}

\newcommand{\De}{\Delta}

\newcommand{\ga}{\gamma}
\newcommand{\Ga}{\Gamma}

\newcommand{\al}{\alpha}

\newcommand{\de}{\delta}
\newcommand{\lga}{\longrightarrow}

\newcommand{\ov}{\over}

\newcommand{\sq}{\sqrt}

\newcommand{\ev}{\equiv}
\newcommand{\lb}{\label}

\newcommand{\PL}[1]{ {\it Phys.~Lett.} {\bf #1}}

\newcommand{\PR}[1]{ {\it Phys.~Rev.} {\bf #1}}
\newcommand{\PRL}[1]{ {\it Phys.~Rev.~Lett.} {\bf #1}}

\newcommand{\JP}[1]{ {\it J.~Phys.} {\bf #1}:\  Math.~Gen.~}

\newcommand{\RMP}[1]{ {\it Rev.~Mod.~Phys.} {\bf #1}}

\newcommand{\JMP}[1]{ {\it J. Math.~Phys.} {\bf #1}}

\newcommand{\CMP}[1]{ {\it Commun.~Math.~Phys.} {\bf #1}}

\begin{document}
\begin{titlepage}
\setcounter{page}{1}
\renewcommand{\thefootnote}{\fnsymbol{footnote}}

\begin{flushright}
hep-th/0109028 
\end{flushright}

\vspace{6mm}
\begin{center}
{\Large\bf Coherent States for Generalized Laguerre Functions} 
\vspace{8mm}

{\large\bf Ahmed Jellal}
\footnote{{\textsf{jellal@gursey.gov.tr --- jellal@na.infn.it}}}

\vspace{5mm}
{\em Feza G\"ursey Institute, P.O. Box 6, 81220
\c{C}engelk\"oy, Istanbul, Turkey}\\ 

\end{center}

\vspace{5mm}
\begin{abstract}
We explicitly construct a Hamiltonian 
whose exact eigenfunctions are the generalized 
Laguerre functions. Moreover, we present
the related raising and lowering
operators. 
We investigate the
corresponding coherent states by adopting
the Gazeau-Klauder approach, where 
resolution of unity
and overlapping properties  
are examined. Coherent states are found to be 
similar to those found for a particle
trapped in a P\"oschl-Teller potential of the trigonometric
type. Some comparisons with 
Barut-Girardello and Klauder-Perelomov
methods are noticed.
\end{abstract}
\end{titlepage}

\newpage

\section{Introduction}
Coherent states were first investigated 
by Schr\"odinger in $1926$ \cite{sch}, where 
he introduced harmonic oscillator coherent states. 
Coherent states are mathematical tools which provide
a close connection between classical and quantum formalisms.
In fact, there appeared many applications of them 
\cite{mal,fel,tam,var} and recently they were used to
study orbital magnetism of two-dimensional electrons \cite{jel1}
and noncommutative magnetism \cite{jel2,jel3}.
 
In $1971$, Barut and Girardello \cite{bg} proposed a method
for constructing coherent states. They defined them as 
eigenstates of the lowering operator of the system.
A generalization of their approach was suggested by
Gazeau and Klauder \cite{gaz2}. Klauder and 
Perelomov, separately, proposed another
definition \cite{kla,per}, which is actually 
known as the Klauder-Perelomov approach. In the latter
coherent states are defined as the states
generated by the action of the elements
of the related dynamical symmetry group on the
Hilbert space whose basis vectors are some
special functions.

On the other hand, special functions
are often investigated by using 
the factorization method. This method 
consists of constructing raising  
and lowering operators which generate
orthogonal bases in terms of
special functions \cite{inf,mil}.

Motivated by the recent developments concerning
special functions \cite{ata,lor},
we consider a Hilbert space whose elements
are generalized Laguerre functions. 
By constructing raising and lowering
operators acting on
these states one can obtain an explicit realization
of the Hamiltonian which is defined to be 
diagonal in this Hilbert space.
We deal with the coherent states of this system in terms
of the Gazeau-Klauder method. The states tourn out
to be similar to those found for a particle
trapped in a P\"oschl-Teller potential of the trigonometric
type. Moreover, we compare the obtained Gazeau-Klauder
coherent states to the Barut-Girardello
and Klauder-Perelomov ones.

In section $2$
we review some properties of generalized Laguerre functions.
In terms of the differential equation 
and the recurrence relations
satisfied by these functions
we construct operators which are acting as 
raising and lowering operators on them.
Then, a Hamiltonian is defined such that generalized
Laguerre functions are its eigenfunctions.
In section $3$ we construct coherent states of this
system following the approach of
Barut-Girardello. 
Section $4$ is devoted to building coherent states
in terms of the method given by Gazeau-Klauder. 
In section $5$, we first give an explicit realization
of the dynamical symmetry group $su(1,1)$. Then, we use
elements of this group to generate coherent states
from the Klauder-Perelomov definition.
We present conclusions and some proposals in the final section.
 
\section{Hamiltonian formalisms}
We start by reviewing some properties
related to the associated Laguerre polynomials 
$L_n^{\al}(x)$ \cite{gra}, which will be used. By definition
$L_n^{\al}(x)$ are 
\beq
L_n^{\al}(x)={1\ov n!}e^{x}x^{-\al}{d^n\ov dx^n}(e^{-x}x^{n+\al})
=\sum_{m=0}^{n}(-1)^m \pmatrix{n+\al\cr n-m\cr}{x^m\ov m!},
\lb{lp}
\eeq
where $\pmatrix{p\cr n\cr}={p(p-1)...(p-n+1)\ov 1.2.3...n},
\pmatrix{p\cr 0\cr}=1$ and $L_n^{0}(x)=L_n(x)$. 
The generating function corresponding to associated 
Laguerre polynomials is 
\beq
\sum_{n=0}^{\infty}L_n^{\al}(x)x^{\al}=
{e^{xz\ov z-1}\ov(1-z)^{\al+1}}.\lb{gf}
\eeq
Note that $L_n^{\al}(x)$ are orthogonal with respect to 
the following weight function
\beq
\rho(x)=x^{\al}e^{-x}, \qquad \al>-1,\lb{wf}
\eeq
satisfy the differential equation
\beq
\Big[x{d^2\ov dx^2}+ (\al-x+1){d\ov dx}+n\Big]
L_n^{\al}(x)=0,\lb{lpde}   
\eeq
and the recurrence relations
\beq
\bra{l}
(n+1)L_{n+1}^{\al}(x)-(2n+\al+1-x)L_{n}^{\al}(x)
+(n+\al)L_{n-1}^{\al}(x)=0,\lb{lprr}\\
x{d\ov dx}L_n^{\al}(x)=nL_n^{\al}(x)-(n+\al)L_{n-1}^{\al}(x). 
\era
\eeq

In terms of $L_n^{\al}(x)$ one can define 
the generalized Laguerre functions as \cite{lor}  
\beq
\psi_n^{\al}(x)=\sq{n!x^{\al+1}e^{-x}\ov\Ga(n+\al+1)}L_n^{\al}(x),
\qquad \al>-1,\lb{glf}
\eeq 
where $\Ga(n+\al+1)$ is the Gamma function
\beq
\Ga(n+\al+1)=\int_{0}^{\infty}e^{-t}t^{n+\al}dt,\lb{gf}
\eeq
with $\Ga(n)=(n-1)!$ The generalized Laguerre functions
$\psi_n^{\al}(x)$ can be shown to obey
the orthonormality condition
\beq
\int_0^{\infty}\psi_n^{\al}(x)\psi_{n'}^{\al}(x)x^{-1}dx=
\de_{nn'}.\lb{oc}
\eeq 
Using the differential equation (\ref{lpde}) and 
the recurrence relations (\ref{lprr}) of
associated Laguerre polynomials,
it is easy to derive the differential equation
\beq
\Big[x{d^2\ov dx^2}+ {1\ov 4}(2\al+2-x+{1-\al^2\ov x})+n\Big]
\psi_n^{\al}(x)=0,\lb{glde}
\eeq
and the recurrence relation 
\beq
\sq{(n+\al+1)(n+1)}\psi_{n+1}^{\al}(x)+\sq{(n+\al)n}\psi_{n-1}^{\al}(x)
-(2n+\al+1-x)\psi_{n}^{\al}(x)=0,\lb{glrr}
\eeq 
of the genaralized Laguerre functions.

By exploring the above formulas, we can define the raising 
operator $A^+$ and the lowering
operator $A^-$ for the generalized Laguerre functions: 
\beq
\bra{l}
A^+=-x{d\ov dx}-{1\ov 2}(2n+\al+1-x),\\
A^-=x{d\ov dx}-{1\ov 2}(2n+\al+1-x).\lb{rl}
\era
\eeq
They act on $\psi_{n}^{\al}(x)$ as follows
\beq
\bra{l}
A^+\psi_{n}^{\al}(x)=-\sq{(n+1)(n+\al+1)}\psi_{n+1}^{\al}(x),\\
A^-\psi_{n}^{\al}(x)=-\sq{n(n+\al)}\psi_{n-1}^{\al}(x).\lb{rla}
\era
\eeq 
$\psi_{n}^{\al}(x)$ can be written in terms of
$A^+$ and $\psi_{0}^{\al}(x)$:  
\beq
\bra{l}
\psi_{n}^{\al}(x)={1\ov\sq{n!(\al+1)_n}}(A^+)^n\psi_{0}^{\al}(x), \\ 
\psi_{0}^{\al}(x)={1\ov\sq{\Ga(\al+1)}}x^{\al+1\ov 2}
e^{-{x\ov 2}},\lb{frg}
\era
\eeq 
where the shifted factorial is 
$(a)_n={\Ga(a+n)\ov\Ga(a)}=a(a+1)...(a+n-1)$. Observe that the 
following relation are satisfied
\beq
\bra{l}
A^+A^-\psi_{n}^{\al}(x)=n(n+\al)\psi_{n}^{\al}(x),\\ 
A^-A^+\psi_{n}^{\al}(x)=(n+1)(n+\al+1)\psi_{n}^{\al}(x).\lb{rlaf}
\era
\eeq 
One can define a Hamiltonian $H$ in terms of raising and lowering 
operators (\ref{rl}) in such a way that
\beq
H=A^+A^-,\lb{hglf}
\eeq     
where the generalized Laguerre functions (\ref{glf}) satisfy
the eigenvalue equations
\beq
H\psi_{n}^{\al}(x)=e_n\psi_{n}^{\al}(x),\lb{ha}
\eeq
with
\beq
e_n=n(n+\al),\qquad n=0,1,2 ...\lb{ev1}
\eeq

Note that the obtained spectrum 
(\ref{ev1}) is similar to that found for a particle
trapped in a P\"oschl-Teller potential of the trigonometric
type. We remind the reader that the coherent states for this system
are constructed by using the Gazeau-Klauder method
\cite{gaz1},  as well as other methods~\cite{nie,dao} . 
   
\section{Barut-Girardello coherent states} 
According to
Barut-Girardello definition, coherent states are 
defined to be the
eigenvalues of lowering operator:
\beq
A^-|z,\al>=z|z,\al>,\lb{bg1}
\eeq
where $z\in\BC$. $|z,\al>$ can be written in terms of the generalized
Laguerre functions (\ref{glf}) as
\beq
|z,\al>=N(z)^{-1}\sum_{n=0}^{\infty}
{z^{n}\ov\sq{n!(\al+1)_n}}|\psi_n^{\al}>,\lb{bg2}
\eeq
where $N(z)$ is the normalization factor
\beq
N(z)={\sq{\Ga(\al+1)I_{\al}(2|z|)}\ov |z|^{\al\ov 2}}.\lb{bgnf}
\eeq
The Barut-Girardello coherent
states (\ref{bg2}) become
\beq
|z,\al>= {|z|^{\al\ov 2}\ov\sq{I_{\al}(2|z|)}}\sum_{n=0}^{\infty}  
{z^{n}\ov\sq{n!\Ga(n+\al+1)}}|\psi_n^{\al}>.\lb{bg3}
\eeq
After some computations, we simplify
the last equation to
\beq
|z,\al>=\sq{x^{-\al}\Ga(\al+1)\ov I_{\al}(2|z|)}
e^{z}J_{\al}(2\sq{xz})|\psi_{0}^{\al}>,\lb{bg4}   
\eeq
where $J_{\al}(2\sq{xz})$ is the Bessel function
\beq
J_{\al}(2\sq{xz})=(xz)^{\al\ov 2}\sum_{n=0}^{\infty}
{(-1)^{n}\ov\sq{n!\Ga(n+\al+1)}}(xz)^n.\lb{bf}
\eeq
We can see that the overlapping of two coherent
states does not vanish
\beq
<z_1,\al|z_2,\al>={I_{\al}(2\sq{{\bar z}_1z_2})\ov
\sq{I_{\al}(2|z_1|)I_{\al}(2|z_2|)}}.\lb{bgop}
\eeq
By the choice of the measure:   
\beq
d\mu(z,\al)={2\ov\pi}K_{\al}(2|z|)I_{\al}(2|z|)d^2z,\lb{bgm}
\eeq
one can show that the resolution of unity is satisfied
\beq
\int |z,\al><z,\al|d\mu(z,\al)=1.\lb{rubg}
\eeq
As a consequence, for any state
$|\Psi>=\sum_{n=0}^{\infty}c_n|\psi_n^{\al}>$
in the Hilbert space, one can construct the analytic
function
\beq
f(z)={N(z)\ov\sq{\Ga(\al+1)}}<z,\al|\Psi>
=\sum_{n=0}^{\infty}{c_n\ov\sq{n!\Ga(n+\al+1)}}z^n.\lb{bf}
\eeq
Therefore, the state $|\Psi>$ can be expressed in terms
of the Barut-Girardello coherent states (\ref{bg4}) 
in such a way that
\beq
|\Psi>=\int d\mu(z,\al){{\bar z}^{\al\ov 2}\ov  
\sq{I_{\al}(2|z|)}}f(z)|z,\al>,\lb{st}
\eeq
and we have
\beq
<\Psi|\Psi>=\int d\mu(z,\al){|z|^{\al}\ov I_{\al}(2|z|)}  
|f(z)|^2<\infty.\lb{ocst}
\eeq

Barut-Girardello coherent states also
have been considered by Brif \cite{bri} in the framework of
the Lie algebra $su(1,1)$. He constructed
coherent states as 
eigenstates of the lowering operator of
$su(1,1)$. We should also mention reference \cite{dod},
where the Barut-Girardello coherent states are constructed.

\section{Gazeau-Klauder coherent states}
Consider a Hamiltonian $H$
with discrete spectrum $(e_n)$ which 
is bounded below and has been
adjusted so that $H\ge 0$. Moreover,
assume that the eigenvalues are 
nondegenerate and arranged in increasing order. 
With these assumptions 
Gazeau and Klauder 
suggested a method where the coherent states 
are characterized by
a real two-parameter set 
$\{|J,\ga>~,\;\;\; J\ge 0,-\infty < \ga < +\infty\}$, such that
\beq
|J,\ga>=N(J)^{-1}\sum_{n=0}^{\infty}{J^{n\ov 2}\ov\sq{\rho_n}}
e^{-ie_n\ga}|\psi_n^{\al}>.\lb{cs1}
\eeq
The positive constants $\rho_n$ are defined by
\beq
\rho_n=e_1e_2... e_n,\lb{rho1}
\eeq
and the normalization factor $N(J)$ 
\beq
N(J)^{2}=\sum_{n=0}^{\infty}{J^{n}\ov\rho_n}.\lb{nf1}
\eeq

We would like to apply this approach to derive 
the corresponding coherent states for the generalized 
Laguerre functions (\ref{glf}). For this purpose, we
start by noting that the energy
spectrum (\ref{ev1}) is 
arranged 
in the strictly increasing order:
\beq
0=e_0<e_1<e_2 ... <e_n< ...\lb{sispn}.
\eeq
Therefore, by inserting (\ref{ev1}) into (\ref{rho1}),
we find
\beq
\rho_n=n!(\al+1)_n,\lb{rho2}
\eeq 
as well as
\beq
N(J)^{2}={\Ga(\al+1)\ov J^{\al\ov 2}}I_{\al}(2\sq{J}),\lb{nf2}
\eeq
where $I_{\al}(2\sq{J})$ are the modified Bessel functions 
\beq
I_{\al}(2\sq{J})=\sum_{n=0}^{\infty}
{J^{n+{\al\ov 2}}\ov n!\Ga(n+\al+1)}.\lb{mbf}
\eeq
In Gazeau-Klauder approach, one needs
to specify the radius of convergence \cite{gaz2}, which
plays an important role in investigating the resolution
of unity. It is defined by
\beq
R=\lim_{n\lga\infty}{\sq[n]{\rho_n}}.\lb{rc1}
\eeq
Then from (\ref{rho2}), we observe that 
\beq
R=\lim\sup_{n\lga\infty}{\sq[n]{n!(\al+1)_n}}=\infty.\lb{rc2}
\eeq
In the present approach, the positive constants $\rho_n$ 
are assumed to arise as the moments of a probability distribution
\beq
\rho_n=n!(\al+1)_n=\int_{0}^{R=\infty}J^n\rho(J)dJ,\lb{rho3}
\eeq
which leads to the following expression for $\rho(J)$
\beq
\rho(J)={2\ov\Ga(\al+1)}J^{\al\ov 2}K_{\al}(2\sq{J}).\lb{rho4}
\eeq  
$K_{\al}(2\sq{J})$ are the $\al-$order modified Bessel
functions of the second kind
\beq
K_{\al}(2\sq{J})={\pi\ov 2\sin(\pi\al)}
(I_{-\al}(2\sq{J})-I_{\al}(2\sq{J})).\lb{mbfia}
\eeq
Actually, (\ref{cs1}) becomes
\beq
|J,\ga>=J^{\al\ov 2}I_{\al}(2\sq{J})\psi_0^{\al}
\sum_{n=0}^{\infty}{J^{n\ov 2}\ov\Ga(n+\al+1)}e^{-in(n+\al)\ga}
|L_n^{\al}>.\lb{cs2}
\eeq  

To complete the construction of coherent states $|J,\ga>$, 
we need to check some requirements. For this, 
we consider the relation
\beq
\int |J,\ga><J,\ga|d\mu(J,\ga)=\lim_{T\lga\infty}{1\ov 2T}\int_{-T}^Td\ga
\Big[\int_0^{\infty} k(J)|J,\ga><J,\ga|dJ\Big],\lb{ru1}
\eeq
where $k(J)$ is defined by \cite{gaz2}
\beq
k(J)=\left\{
\bra{l}
\{N(J)^2\rho(J)\ge 0,\qquad 0\le J<R,\\
\rho(J)\ev 0,\qquad\qquad J>R.
\era
\right.\lb{kj1}        
\eeq
According to our data, $k(J)$ is nothing but
\beq
k(J)=2I_{\al}(2\sq{J})K_{\al}(2\sq{J})\ev k_{\al}(J) .\lb{kj2}
\eeq   
Now it is easy to observe that the resolution of unity is satisfied
\beq
\int |J,\ga><J,\ga|d\mu(J,\ga)=1,\lb{ru2}
\eeq
where $\ga\in [-\pi,\pi]$. The temporal stability is immediate
\beq
e^{-iHt}|J,\ga>=|J,\ga+t>.\lb{ts}
\eeq
The action of identity is given by
\beq
<J,\ga|H|J,\ga>=N(J)^{-2}\sum_{n=0}^{\infty}{n(n+\al)\ov 
n!(\al+1)_n}J^{n}.\lb{ai1}
\eeq 
Using (\ref{nf2}) and making some calculations, we find
\beq
<J,\ga|H|J,\ga>=J.\lb{ai2}
\eeq  
The overlapping is
\beq
<J',\ga'|J,\ga>={2\ov N(J)N(J')}\sum_{n=0}^{\infty}
{(JJ')^{n\ov 2}\ov n!(\al+1)_n}e^{-ie_n(\ga-\ga')}.\lb{ov1}
\eeq 
It is clear from the last equation that two different coherent states
are not orthogonal to each other. If
we made a restriction such that $\ga=\ga'$, then
the overlapping becomes
\beq
<J',\ga|J,\ga>={I_{\al}(2\sq[4]{JJ'})\ov
\sq{I_{\al}(2\sq{J'})I_{\al}(2\sq{J'})}}.\lb{ov2}
\eeq 

We conclude that the obtained coherent states 
(\ref{cs2}) 
are similar to 
those found for particle moving in 
the P\"oschl-Teller potential \cite{gaz1}
by adopting Gazeau-Klauder approach. 
Given this similarity,
one can translate directly the physical 
interpretation
of P\"oschl-Teller coherent states to
generalized Laguerre ones. We 
remind our readers that the authors of \cite{gaz1} 
have studied the
spatial and temporal features of the mentioned
coherent states. 

We are going to discuss
the weighting distribution $|c_n|^2$ corresponding
to our coherent states (\ref{cs2}), which can
be written in terms of the $J$ parameter
\beq
|c_n|^2={J^n\ov N(J)^2\rho_n}.\lb{wp}
\eeq 
Furthermore, there is a parameter called Mandel parameter
$Q$ which
plays an important role, since it can determine the
nature of the weighting distribution $|c_n|^2$ as we will
show. It is 
defined by \cite{man,sol}
\beq
Q={(\De n)^2\ov <n>}-1,\lb{man1}
\eeq
where $<n>$ are the mean values 
\beq
<n>=\sum_{n=0}^{\infty}n{J^2\ov N(J)^2\rho_n},\lb{mvs1}\\
\eeq
and the spread is
\beq
\De n=[<n^2>-<n>^2]^{1\ov 2}.\lb{spread1}
\eeq   
Note that, $Q=0$ yields $(\De n)^2=<n>$. Thus,
for $Q=0$ 
the weighting distribution 
becomes to be Poissonian
\beq
|c_n|^2={1\ov n!}<n>^ne^{-<n>}.\lb{poi}
\eeq
Otherwise, it is 
super-Poissonian
or sub-Poissonian for $Q$ 
strictly positive or negative.
To determine explicitly the nature of
$|c_n|^2$ in our case, we need to evaluate
(\ref{mvs1}) and (\ref{spread1}). 
A direct calculation leads to
\beq
<n>=\sq{J}{I_{\al+1}(2\sq{J})\ov I_{\al}(2\sq{J})},\lb{mvs2}\\
\eeq
and
\beq
\De n=\Big[<n>-<n>^2+J{I_{\al+2}(2\sq{J})\ov 
I_{\al}(2\sq{J})}\Big]^{1\ov 2}.\lb{spread2}
\eeq  
Then (\ref{man1}) becomes
\beq
Q=\sq{J}\Big[{I_{\al+2}(2\sq{J})I_{\al}(2\sq{J})
-(I_{\al+1}(2\sq{J}))^2
\ov I_{\al+1}(2\sq{J})I_{\al}(2\sq{J})}\Big].\lb{man2}
\eeq
Since $(I_{\al+1}(2\sq{J}))^2\ge
I_{\al+2}(2\sq{J})I_{\al}(2\sq{J})$, we realize 
immediately that we have a negative
$Q$, which implies that the weighting 
distribution (\ref{wp}) 
is sub-Poissonian. The Poissonian
case can be recovered when $J$ is large 
(for more details see \cite{gaz1}). 

We close this section by 
noting that the obtained overlapping property 
for Gazeau-Klauder approach 
at $\ga=\ga'$ 
(\ref{ov2}) is similar to
that derived from Barut-Girardello definition 
(\ref{bgop}) although, the
coherent states 
(\ref{bg4}) and (\ref{cs2})
are not the same. However, 
by choosing $\ga=0$ and $J$ as a complex
parameter, we can reproduce the Barut-Girardello coherent
states (\ref{bg4}) from the 
Gazeau-Klauder ones (\ref{cs2}).  

\section{Klauder-Perelomov coherent states} 
Following the Klauder-Perelomov definition, 
coherent states of a given system can
be constructed in terms of its dynamical symmetry
group. Thus, 
we need first to determine the appropriate dynamical symmetry group
of the system described by the generalized Laguerre functions
(\ref{glf}). 
Starting from (\ref{rlaf}), one can show that the following 
commutation relation is satisfied
\beq
[A^-,A^+]\psi_{n}^{\al}(x)=(2n+\al+1)\psi_{n}^{\al}(x).\lb{rlc}
\eeq
Let us introduce the operator $A_3$ defined to satisfy
\beq
A_3\psi_{n}^{\al}(x)={1\ov 2}(2n+\al+1)\psi_{n}^{\al}(x).\lb{do}
\eeq
Now the operators $A^-,A^+$ 
and $A_3$ generate the $su(1,1)$ Lie algebra
\beq
[A^-,A^+]=2A_3,\qquad [A_3,A^+]=A^+,\qquad [A_3,A^-]=-A^-.\lb{la1}
\eeq
The corresponding Casimir operator is 
\beq
C=A_3^2-{1\ov 2}(A^+A^-+A^-A^+).\lb{casi}
\eeq
The value of $C$ in the Hilbert space of the
generalized Laguerre functions (\ref{glf}) is
\beq
{1\ov 4}(\al+1)(\al-1).\lb{casa}
\eeq
This means that the unitary irreducible 
representations of $su(1,1)$ are determined by 
the $\al$-parameter.

The Klauder-Perelomov definition
of coherent states consists of applying the operator 
$e^{\xi A^+}$ on the ground state $\psi_0^{\al}$, such that
\beq
|\xi,\al>=e^{\xi A^+-{\bar \xi}A_{-}}|\psi_0^{\al}>,\lb{pcs1}
\eeq 
which leads to
\beq
|z,\al>=(1-|z|^2)^{\al+1}e^{z A^+}|\psi_0^{\al}>,\lb{pcs2}
\eeq  
where $z={\xi\ov |\xi|}\tanh|\xi|$ is a complex number
satisfying the condition $|z|<1$.
The last equation can be reorganized as follows
\beq
|z,\al>=(1-|z|^2)^{\al+1}\sum_{n=0}^{\infty}
\sq{(\al+1)_n\ov n!}z^{n}|\psi_n^{\al}>.\lb{pcs3}
\eeq
Finally, we find that
\beq
|z,\al>=({1-|z|^2\ov 1-z})^{\al+1}e^{xz\ov z-1}
|\psi_0^{\al}>.\lb{pcs4}
\eeq
The overlapping property is
\beq
<z_1,\al|z_2,\al>=[(1-|z_1|^2)(1-|z_2|^2)]^{\al+1\ov 2}
(1-z_1{\bar z}_2)]^{-\al-1},\lb{op}
\eeq  
which shows that the $su(1,1)$ coherent states are normalized,
but are not orthogonal to each other. The resolution of unity
can be obtained with an appropriate choice of the measure. By
choosing it as
\beq
d\mu(z,\al)={\al\ov\pi}{d^2z\ov (1-|z|^2)^2},\lb{m}
\eeq
we get
\beq
\int |z,\al><z,\al|d\mu(z,\al)=1.\lb{pru}
\eeq
As noted for Barut-Girardello coherent states, for any state
$|\Psi>=\sum_{n=0}^{\infty}c_n|\psi_n^{\al}>$
in the Hilbert space, one can construct an analytic
function 
\beq
f(z)= (1-|z|^2)^{-{\al+1\ov 2}}<z,\al|\Psi>=\sum_{n=0}^{\infty}
c_n \sq{(\al+1)_n\ov n!}({\bar z})^n.\lb{pbf}
\eeq
In terms of the $su(1,1)$ coherent states (\ref{pcs4}) 
$|\Psi>$ can be written as 
\beq
|\Psi>=\int d\mu(z,\al)(1-|z|^2)^{{\al+1\ov 2}}
f(z)|z,\al>,\lb{pst}
\eeq
leading to
\beq
<\Psi|\Psi>=\int d\mu(z,\al)(1-|z|^2)^{\al+1}
|f(z)|^2<\infty.\lb{pocst} 
\eeq

Note that the coherent states
derived in the third approach (\ref{pcs4}) are completely
different from the previous ones,
namely the Barut-Girardello
(\ref{bg4}) and the Gazeau-Klauder
(\ref{cs2}) coherent states. Moreover,
their overlapping property is also different from
the others (\ref{bgop})-(\ref{ov1}).

We close this section by noting
that Klauder-Perelomov coherent states 
have been studied by Trivinov in the context of more physical
problem of the "singular oscillator" described in terms of the
Hamiltonian $p^2 + w^2(t) x^2 + g/x^2$ \cite{tri1}.

\section{Conclusions}
Generalized Laguerre functions are considered
as a basis of a Hilbert space. 
Raising, lowering 
and Hamiltonian operators are constructed. 
The corresponding 
coherent states are inevestigated
by using three different methods. 
The resolution of unity and the overlapping properties 
have been considered in each case. We found that 
although one can
recover the Barut-Girardello coherent states (\ref{bg4}) 
from the Gazeau-Klauder ones (\ref{cs2}) under the
conditions: $\ga=0$ and $J$ is a complex
parameter, they differ from the coherent states (\ref{pcs4}) 
obtained in terms of the Klauder-Perelomov apparoach. 

Obviously, other special functions can be
studied by the same method presented here. Moreover,
one could examine  
the states minimizing the Robertson-Schr\"odinger
uncertainty relation (intelligent states)
\cite{tri2} in terms of
the generalized Laguerre functions~(\ref{glf}). 

\section*{Acknowledgements}
The author would like to express his gratitude to Professor
\"O.F. Dayi for valuable conversations
and for his advice. He is indebted to Prof. 
C. Saclioglu for a careful reading of the
manuscript.
He is very
thankful to Professors G. Ortiz and T. Palev for their
helps and encouragments. He is also 
grateful to Professor E. In\"on\"u for his
interest and encouragment.

\end{document}